\documentclass[aps,twocolumn,showpacs,superscriptaddress]{revtex4}
\usepackage{epsfig}
\usepackage{times}
\usepackage{amsmath}
\begin{document}

\title{Evolutionary games on multilayer networks: A colloquium}

\author{Zhen Wang}
\email{zhenwang0@gmail.com}
\affiliation{School of Computer and Information Science, Southwest University, Chongqing, 400715, China}
\affiliation{Department of Physics, Hong Kong Baptist University, Kowloon Tong, Hong Kong}

\author{Lin Wang}
\email{fdlwang@gmail.com}
\affiliation{Centre for Chaos and Complex Networks, City University of Hong Kong, Kowloon Tong, Hong Kong}

\author{Attila Szolnoki}
\email{szolnoki@mfa.kfki.hu}
\affiliation{Institute of Technical Physics and Materials Science, Centre for Energy Research, Hungarian Academy of Sciences, P.O. Box 49, H-1525 Budapest, Hungary}

\author{Matja{\v z} Perc}
\email{matjaz.perc@uni-mb.si}
\affiliation{Faculty of Natural Sciences and Mathematics, University of Maribor, Koro{\v s}ka cesta 160, SI-2000 Maribor, Slovenia}
\affiliation{Department of Physics, Faculty of Sciences, King Abdulaziz University, Jeddah, Saudi Arabia}
\affiliation{CAMTP -- Center for Applied Mathematics and Theoretical Physics, University of Maribor, Krekova 2, SI-2000 Maribor, Slovenia}

\begin{abstract}
Networks form the backbone of many complex systems, ranging from the Internet to human societies. Accordingly, not only is the range of our interactions limited and thus best described and modeled by networks, it is also a fact that the networks that are an integral part of such models are often interdependent or even interconnected. Networks of networks or multilayer networks are therefore a more apt description of social systems. This colloquium is devoted to evolutionary games on multilayer networks, and in particular to the evolution of cooperation as one of the main pillars of modern human societies. We first give an overview of the most significant conceptual differences between single-layer and multilayer networks, and we provide basic definitions and a classification of the most commonly used terms. Subsequently, we review fascinating and counterintuitive evolutionary outcomes that emerge due to different types of interdependencies between otherwise independent populations. The focus is on coupling through the utilities of players, through the flow of information, as well as through the popularity of different strategies on different network layers. The colloquium highlights the importance of pattern formation and collective behavior for the promotion of cooperation under adverse conditions, as well as the synergies between network science and evolutionary game theory.
\end{abstract}

\maketitle

\section{Introduction}
The hallmark property of a complex system is that a large number of simple units give rise to fascinating collective phenomena that could not be anticipated from an individual unit \cite{gellmann_es88}. Social order, biological complexity, brain power, ant colonies, and economic interconnectedness are all prime examples of topics one might attempt to study with a complex system at the heart of the research endeavor. But what is behind the emergent complexity? What turns people to societies and simple cells like neurons to a brain? The answer is, primarily, the network. Although phenomena such as self-organization and pattern formation might play a pivotal role too, it is mainly the way the simple units that form the complex system are connected with each other that makes them so much more than just the sum of their parts. Recent decades have seen an unprecedented development of data-driven mathematical models that offer fresh new insights into complex systems, culminating into a new discipline named network science \cite{barabasi_np12}.

Despite its youth, network science is enjoying widespread recognition and appeal because many natural and social systems owe their functionality to a complex network as their backbone \cite{boccaletti_pr06}. Here nodes are the units or components that make up the system, while the links among them determine who interacts with whom. Pioneering works \cite{watts_dj_n98, barabasi_s99} have identified surprising similarities in networks describing very different natural and man-made systems, such as short average path lengths between pairs of nodes paired with a relatively high clustering of node triples, or heavy tailed distributions of node degree. These universalities have propelled network science to become one of the hottest research disciplines in the 21st century \cite{boccaletti_pr06, albert_rmp02, newman_siamr03}.

Almost simultaneously with network science, in evolutionary game theory \cite{maynard_82, hofbauer_98, nowak_06, sigmund_10} leaps of progress have also been made, largely due to interdisciplinary approaches that link together knowledge from biology, sociology, economics as well as mathematics, physics and psychology \cite{weibull_95, szabo_pr07, schuster_jbp08, perc_bs10, frey_pa10, roca_plr09, santos_jtb12, rand_tcs13, perc_jrsi13, szolnoki_jrsif14}. Evolutionary games provide a comprehensive theoretical framework to investigate strategic choices in a broad variety of complex systems \cite{ichinose2013adaptive, fu2009partner, chen2008interaction, ichinose2013robustness, li2014comprehensive, chen2007prisoner, xia2012effects, xia2011enhancement, wang2008phase, wu2009effects, wu2009diversity, leeprl11, cui14, rong2010emergence}. Based on the fundamentals of evolutionary game theory, decision-making has been extensively applied to the fields of species variety \cite{reichenbach2007mobility}, climate negotiation \cite{milinski2008collective, pacheco_plrev14}, public health \cite{bauch2004vaccination, zxwu13} as well as traffic flow \cite{tanimoto2014dangerous}, to name but a few examples.

Despite numerous practical ramifications and application areas mentioned above, however, the main fundamental problem that is studied in the realm of evolutionary game theory is the evolution of cooperation \cite{axelrod_84,nowak_11}. Cooperation is an altruistic act that is costly to perform but benefits others. Eusocial insects like ants and bees are famous for their large-scale cooperative behavior \cite{wilson_71,wang2008trade}. Cooperation is also found in birds, where helpers often take care for the offspring of others \cite{skutch_co61}. Humans have recently been dubbed supercooperators \cite{nowak_11} for our unparalleled other-regarding abilities and cooperative drive. Importantly, altruistic cooperation is the most important challenge to Darwin's theory of evolution, and it is fundamental for the understanding of the main evolutionary transitions that led from single-cell organisms to complex animal and human societies \cite{maynard_95,pennisi_s05}. As such, understanding the evolution of cooperation remains a grand challenge that continues to attract research across the social and natural sciences. Although studies in evolutionary game theory have already revealed fundamental rules that promote cooperation \cite{axelrod_84,nowak_s06}, the chasm behind the Darwinian ``only the fittest survive'' and the abundance of cooperation in human and animal societies remains quite overwhelming.

Interestingly, the relevance of networks for the outcome of evolutionary games has been recognized already in the early 90s, when Nowak and May discovered network reciprocity \cite{nowak_n92b}. More precisely, they have observed that on a lattice cooperators are able to survive by forming compact clusters, and so protect themselves against the exploitation by defectors even if the governing game is a social dilemma where in a well-mixed population defectors would dominate completely. Remarkably, Rand et al. \cite{rand_pnas14} have recently validated this theoretical prediction in a large-scale human experiment. However, it was not until Santos and Pacheco \cite{santos_prl05} discovered that scale-free networks provide a unifying framework for the evolution of cooperation \cite{santos_pnas06, gomez2007dynamical, szolnoki_pa08} that the field of evolutionary games on network really took off. Since then several excellent works have elaborated on the relevance of the network structure for the evolution of cooperation, as reviewed comprehensively in \cite{szabo_pr07,perc_bs10}. It turned out that not only the structure of the network but also the character of the interactions could be decisive. Namely, multi-point interactions \cite{perc_jrsi13} or cyclically dominant relations among strategies \cite{szolnoki_jrsif14} can further amplify the importance of the population being structured rather than well-mixed.

Most recently, the attention has been shifting away from single, isolated networks to networks of networks, or so-called interdependent or multiplex or multilayer networks \cite{buldyrev2010catastrophic, gao_jx_nsr14, helbing_n13, kivela_jcn14, boccaletti_pr14, d2014networks}. Networks of networks have been brought to the spotlight by the discovery that even small and seemingly irrelevant changes in one network can have catastrophic and very much unexpected consequence in another network \cite{buldyrev2010catastrophic}. Indeed, not only are our interactions limited and thus inadequately described by well-mixed models, it is also a fact that the networks that should be an integral part of such models are often interconnected, thus making the processes that are unfolding on them interdependent. From the World economy and transportation systems to social media, it is clear that processes taking place in one network might significantly affect what is happening in many other networks. Networks of networks are therefore a more apt description of such interdependent systems.

As we hope this colloquium will succeed in demonstrating based on current research, taking into account the fact that humans are typically members in many different social networks has important consequences for the evolution of cooperation. A few simple considerations illustrate the case in point. When we choose a certain strategy, this choice is likely to be perceived differently among our friends, in the workplace, and within our family. It is also likely that the purchase from the strategy is going to be different in these different networks. Alternatively, one can choose a different strategy in each network to try and optimize the outcome, especially if the networks are only weekly interconnected. Either way, it is natural to clarify how such considerations might affect cooperative behavior, and by doing so move towards a more practical and realistic modeling of human cooperation. In what follows, we first review the general framework of multilayer networks, then continue with the review of evolutionary games on multilayer networks, and lastly, we conclude and provide a concise outlook.

\section{From single-layer towards multilayer networks}
Research dealing with or using networks has had, and in fact still very much has, a strong appeal and impact across a myriad of scientific disciplines \cite{boccaletti_pr06, albert_rmp02, newman_siamr03, dorogovtsev2002evolution}. Based on the most basic definition of a network, many real-world entities can be quantitatively described by means of a network \cite{boccaletti_pr06, boccaletti_pr14}. A network typically consists of nodes or vertices that are connected by links or edges. For example, networks have been used to describe interactions between neurons, the trade among markets, the relationship among words, as well as of course the Internet and the World Wide Web \cite{boccaletti_pr06, albert_rmp02, albert1999internet}. Information in the form of rumors, messages, or digital viruses can be transmitted through networks, as well as infectious disease, merchandize, and public goods. Due to the ubiquity of networks, several algorithms have been proposed that describe the most important structural properties of real-world networks \cite{dorogovtsev2002evolution, bornholdt2003handbook}. Prominent examples include the Erd{\H o}s-R{\'e}nyi random network \cite{erd6s1960evolution}, the Watts-Strogatz small-world network \cite{watts_dj_n98}, or the Barab{\'a}si-Albert scale-free network \cite{barabasi_s99}. The rapidly growing availability of human generated data \cite{wasserman1994social}, together with the ever-increasing computational capabilities continue to drive progress in this field, in turn leading to the development of new and more complex theoretical models that are able to accurately describe certain aspects of reality, as reviewed comprehensively in several reviews devoted to networks \cite{boccaletti_pr06, albert_rmp02, newman_siamr03, barthelemy2011spatial, holme2012temporal}.

Despite great achievements during the past couple of decades, traditional research concerning networks assumes that nodes are connected to each other within the same, isolated infrastructure, i.e., the so-called single-layer network. This assumption, however, may in some cases be an oversimplification, given that certain nodes can simultaneously be the building blocks of more than just the one network. And this important consideration applies to natural as well as social systems \cite{de2013mathematical}. For example, major cities are interconnected not just by means of roads, but also by means of rails, as well as by means of air transport. Similarly, people interact face-to-face, via phone, on online social networks, in their work environment, and so on \cite{min2013layer}. It is thus often justified to abandon the traditional assumption of a single-layer network and replace it with a multilayer network formalism. Not surprisingly then, the multilayer network, defined as a combination class of networks that are interrelated in a nontrivial way, has recently emerged as a fundamental concept to quantitatively describe the interactions not just within, but also among different networks \cite{kivela_jcn14, boccaletti_pr14, de2013mathematical}.

With regards to terminology, the term multilayer network is used here to refer to the rather broad variety of network models involving several networks or network layers, including interconnected networks \cite{wang2013effect, aguirre2014synchronization, donges2011investigating, saumell2012epidemic}, interdependent networks \cite{parshani_prl10, huang_xq_pre11, gao_jx_np12, li_w_prl12, dong_g_epl13}, multiplex networks \cite{gomez_prl13}, networks of networks \cite{pocock2012robustness, d2014networks}, as well as multivariate networks \cite{pattison1999logit}. Although the generic term ``multilayer'' can actually be traced back to sociological and engineering problems of the late 1930s, the efforts of developing a theory of multilayer networks as well as the definition of concepts and methods for quantifying their structural properties is of course a matter of current research.

\begin{figure}
\centerline{\epsfig{file=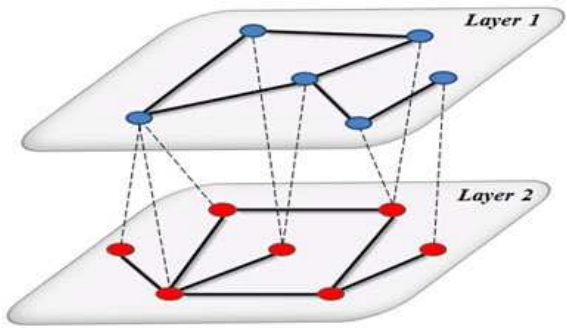,width=7.5cm}}
\caption{Schematic illustration of a multilayer network that is composed of two networks layers. In each layer, nodes have different intra-layer connectivity (solid lines). Moreover, most nodes have an inter-layer link to the corresponding node in the other layer. Exchange of information or similar is thus possible across the two layers.}
\label{illus}
\end{figure}

In particular, the discovery of discontinuous phase transitions that are brought about by cascading failures in interdependent networks \cite{buldyrev2010catastrophic, gao2011robustness, son2011percolation}, where seemingly irrelevant changes in one network can have unexpected and indeed often catastrophic consequences in another network, has lead to an even stronger interest in network science. After this seminal finding, many related works have dealt with multilayer networks, for example related to their robustness against attack and assortativity \cite{zhou_d_pre13, peixoto2012evolution} or their percolation properties \cite{parshani_prl10, bianconi2014multiple}. Likewise, dynamical processes on multilayer networks, diffusion \cite{gomez2013diffusion}, disease spreading and prevention \cite{wang2012dynamics, zhao2014immunization, buono2014epidemics}, evolutionary games \cite{wang_z_epl12, santos2014biased, gomez-gardenes_srep12, jin_q_srep14, wang_b_jsm12, tang_cb_pone14}, voting \cite{diakonova2014absorbing}, and traffic \cite{tan2014traffic}, have all become hot topics of general interest as well.

What sets a multilayer network apart from the traditional single-layer network is that a multilayer network typically consists of $M$ ($M\ge2$) networks (or layers), where the nodes in each network (layer) are connected via intra-layer links, but there are also inter-layer links that link together nodes from other networks. Sometimes the inter-layer links do not serve to connect the nodes, but merely serve to communicate information or some other form of influence between the nodes forming the $M$ networks. Sometimes also the same node appears in more than one network, and sometimes all the nodes pertain to all $M$ networks with the difference between them being the intra-layer links. Depending on these particularities, the terminology that is used also varies. In Fig.~\ref{illus} we provide a schematic illustration of a multilayer network, where, in addition to intra-layer links, most nodes have one or more links to the nodes in the other layer.

\subsection{Basic concepts and definitions of multilayer networks}
In network science, special quantities have been introduced to mathematically determine and analyze the properties of networks. Well-known and widely used examples of such quantities include the degree of a node, betweenness centrality, average path length, clustering coefficient, and the degree distribution to name but a few. And these have been in use for both theoretical and empirical research. In the continuation of this section (in subsequent subsections), we will briefly review how the definitions of some of these quantities have been amended to account for the concept of multilayer networks.

A single-layer network or a graph is usually given in the form $G=(V, E)$, where $V$ is the set of nodes and $E\subseteq V\times V$ corresponds to set of edges that connect said nodes \cite{bollobas1998modern}. It is relatively straightforward to generalize this form to multilayer networks. In particular,
\begin{equation}
G_M=(V_M, E_M),
\label{definition}
\end{equation}
where $M$ is the number of network layers. The set of $E_M$ is the combination of nodes of all the network layers
\begin{equation}
V_M=\cup^{M}_{\alpha=1} V_{\alpha},
\label{node1}
\end{equation}
and
\begin{equation}
V_{\alpha}=\{V^{\alpha}_{1},...,V^{\alpha}_{{N}_{\alpha}}\},
\label{node2}
\end{equation}
where $N_{\alpha}$ is the number of nodes in network layer $\alpha$.

It is worth pointing out that the number of nodes $N_{\alpha}$ can be identical or different in each layer. Every node can have one, more, or even no counterparts in other layers. With respect to the set of connections $E_M$, we can expand further by writing
\begin{equation}
E_M=\{E_{\alpha}\cup E_{\alpha\beta}; \alpha,\beta \in\{1,...,M\}, \alpha\ne\beta\},
\label{edge}
\end{equation}
where $E_{\alpha}\subseteq V_{\alpha}\times V_{\alpha}$ (or $E_{\beta}\subseteq V_{\beta}\times V_{\beta}$) is the set of intra-layer connections in the network layer $\alpha$ (or $\beta$); $E_{\alpha\beta}\subseteq V_{\alpha}\times V_{\beta}$ represents the set of so-called inter-layer connections among network layer $\alpha$ and layer $\beta$. If there exists an intra-connected edge between node $i$ and node $j$ in network layer $\alpha$, the element $a^{\alpha}_{ij}$ of the intra-connection adjacency matrix $A^{\alpha}$ is equal to 1 (namely, $a^{\alpha}_{ij}=1$, $i,j\in(1,...,N_{\alpha})$); otherwise it is equal to 0. Similarly, the element $a^{\alpha\beta}_{ij}$ of the inter-connection adjacency matrix $A^{\alpha\beta}$ is equal to 1 (namely, $a^{\alpha\beta}_{ij}=1$, $i\in(1,...,N_{\alpha})$,  $j\in(1,...,N_{\beta})$) if there is a correlation between node $i$ from network layer $\alpha$ and node $j$ from network layer $\beta$; otherwise $a^{\alpha\beta}_{ij}=0$.

Based on the theory and formalism of traditional single-layer networks \cite{boccaletti_pr06, albert_rmp02, newman_siamr03}, other properties and concepts, like the weight of links, the direction of links, the adaptive nature of networks, can be further incorporated into the multilayer formalism \cite{menichetti2014weighted, shai2013coupled}. We can take adaptive connections as an example, where we just need to consider the adjacency matrices $A_{\alpha}$ and $A_{\alpha\beta}$ as a function of time (namely, $a^{\alpha}_{ij}(t)$ and $a^{\alpha\beta}_{ij}(t)$). While for weighted or directed networks, we simply need to introduce a weight or a direction set $W_M$ into the general framework, such as $G_M=(V_M, E_M,W_M)$. Such additional considerations are usually unavoidable in the analysis of empirical networks, especially given the current outbreak of available data.

\subsection{The structure of multilayer networks}
Because nodes usually play different roles in the structure of the network and in the dynamical processes taking place on the network \cite{boccaletti_pr06, newman_siamr03}, it is necessary, and indeed crucial, to examine the role of nodes in determining the structural properties of a multilayer network. We review specific quantities in the following subsections.

\subsubsection{Node degree and related properties}
In a single-layer network, the degree $k_i$ of a node $i$ is defined as the number of nodes that connect to it. This definition can be naturally extended to the framework of multilayer networks. Thus far, there have been several methods to generalize node degree for multilayer networks, but probably the most common way to do is by means of network aggregation \cite{battiston2014structural}. For example, the degree of node $i$ in a multilayer network can be written with the vector \cite{kivela_jcn14, boccaletti_pr14}
\begin{equation}
{\bf k_i}=(k_{i1},...,k_{iM}),
\label{degree}
\end{equation}
where $k_{i\alpha}$ is the degree of node $i$ in network layer $\alpha$. Equivalently, we can also get the neighborhood, which is the combination of its immediate neighbors in each layer. However, since ${\bf k_i}$ is a vector, it is difficult to obtain a uniform ranking for the degree of nodes. Along this line, related formulations are possible, such as the threshold degree, the multidegree, or the overlapping degree (see \cite{bianconi2013statistical} for details). Moreover, based on the notion of walks and paths, a series of methods measuring the distance between nodes has been proposed as well, first for single-layer and then for multilayer networks \cite{costa2007characterization, nicosia2013growing}.

\subsubsection{Clustering coefficient}
The clustering coefficient is usually used to measure the transitivity of a network. Its value $c_i$ corresponds to the ratio of exiting links to all the possible links among the neighbors of a given node $i$, and the global clustering coefficient $C$ is the average of the clustering coefficients of all the nodes (namely, $C=\sum^N_{i=1}c_i$/N) \cite{boccaletti_pr06, watts_dj_n98, newman_siamr03, bollobas1998modern}. Another alternative definition for the clustering coefficient is the fraction of closed triples among all the possible triads \cite{latora2003economic}. Because of this relative freedom in the definition for isolated networks, the clustering coefficient for a multilayer network potentially gives rise to a whole class of definitions \cite{donges2011investigating, cozzo2013clustering, battiston2013metrics}. There is, however, a generic consideration, which is to involve the average of the value of the clustering coefficient for both inter-layer and intra-layer links. For example, based on the set of neighbors and subgraph projection networks, the authors in \cite{criado2012mathematical} translate the clustering coefficient into a function of each network layer and the projection network. Perhaps most elegantly, borrowing the terminology of 2-triangles and 3-triangles where nodes are located in different layers, the clustering coefficient of a multilayer network can be expressed as the average over clustering coefficient values of all the nodes \cite{battiston2014structural}. Along this line, the clustering coefficient can even be introduced for weighted and directed networks \cite{kivela_jcn14, boccaletti_pr14}.

\subsubsection{Degree-degree correlation}
Degree correlation is another important quantity, which is traditionally used to measure the mixing pattern of nodes in isolated, single-layer networks \cite{boccaletti_pr06, albert_rmp02, newman_siamr03}. If large-degree nodes are more likely to connect to large-degree (small-degree) nodes, the network shows assortative (disassortative) mixing, leading to positive (negative) values of the correlation coefficient $r$ \cite{newman2002assortative, xulvi2004reshuffling}. However, if one tries to translate this quantity directly for use in multilayer networks several considerations are first in order. Evidently, apart from the correlation of nodes in the local layer, a new method is needed to quantify the assortativity or disassortativity correlation between nodes across layers. This was the motivation behind the introduction of the degree-degree correlation coefficient $r_{\alpha\beta}$, which was designed to fill this gap. If large-degree nodes are more inclined to interconnect with large-degree (small-degree) counterparts in the other networks, its value is positive (negative), which shows the assortative (disassortative) mixing pattern between layers. The correlation of network layers is particularly obvious in online social systems. For example, a famous person, a sports hero or a movie star, is likely going to be a hub node in several online networks, like Facebook or Twitter.

\begin{figure*}
\centerline{\epsfig{file=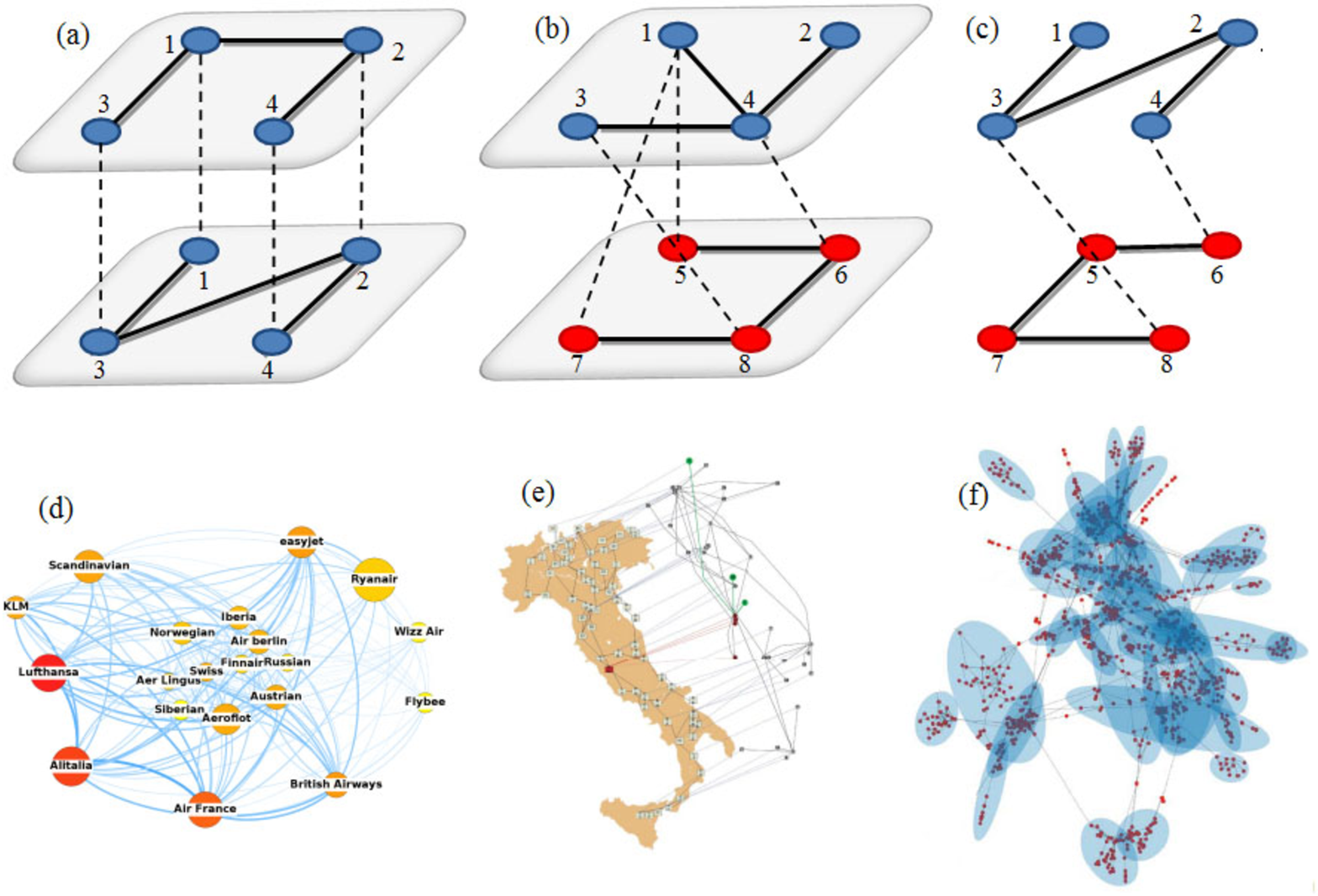,width=14.5cm}}
\caption{Top panel features the schematic illustration of different types of multilayer networks, where solid (dash) lines denote the intra-layer (inter-layer) links. From (a) to (c) they are multiplex networks, interdependent networks, and interconnected networks, respectively. In each scheme, if the nodes of each layer are the same, we use the same color and number to mark them; otherwise a different color and number are used. In addition, inter-layer links for interdependent networks and interconnected networks are different. We note that interdependent networks have dependency links but no actual, physical links across network layers. On the other hand, such across-layer links exist in interconnected networks. Bottom panel features empirically measured, actual multilayer networks. In particular, (d) shows the multiplex network composed by 20 European airline companies (see \cite{nicosia2014measuring}), (e) shows the interdependent networks of the Italian power grid and the computer network (see \cite{buldyrev2010catastrophic}), while (f) shows the interdependent networks formed by the structural subgraphs or communities in a protein-protein interaction network (see \cite{yang2014overlapping}).}
\label{illustrate2}
\end{figure*}

Up to now, several different methods to determine the correlation across network layers have been proposed \cite{bianconi2013statistical, min2014network, watanabe2014cavity, melnik2014dynamics, de2013centrality, battiston2014structural}. One that we single out is the Pearson correlation coefficient, which seems to attract the most interest \cite{battiston2014structural}. It is expressed as
\begin{equation}
r_{\alpha\beta}\equiv\frac{\langle(k_{i\alpha}-\langle k_{\alpha}\rangle)(k_{i\beta}-\langle k_{\beta}\rangle)\rangle}{\sigma_{\alpha}\sigma_{\beta}}=\frac{\langle k_{i\alpha}k_{i\beta}\rangle-\langle k_{i\alpha}\rangle\langle k_{i\beta}\rangle}{\sigma_{\alpha}\sigma_{\beta}},
\label{Pearson}
\end{equation}
where $\langle k_{\alpha}\rangle$ represents the average degree of network layer $\alpha$, and $\sigma_{\alpha}=\sqrt{\langle k_{i\alpha}k_{i\alpha}\rangle-\langle k_{i\alpha}\rangle ^2}$ is the standard deviation of node degree in layer $\alpha$.

Moreover, in \cite{parshani2010inter} the authors define inter degree-degree correlation between a pair of dependent nodes on interdependent networks. If two networks have the degree distribution $p_{k_{\alpha}}$ and $p_{k_{\beta}}$, the correlation level between a pair nodes (one node $i$ from network $\alpha$ with degree $k_{i\alpha}$, another node $j$ from network $\beta$ with degree $k_{i\beta}$) is
\begin{equation}
\varsigma=\sum_{k_{i\alpha} k_{j\beta}} k_{i\alpha} k_{j\beta}(e_{k_{i\alpha}k_{j\beta}}- p_{k_{i\alpha}}p_{k_{j\beta}}),
\label{inter-degree}
\end{equation}
where $e_{k_{i\alpha}k_{j\beta}}$ denotes the joint probability that a dependency link connects both nodes. Normalizing by the maximum value $\varsigma_{max}$, a general measure
\begin{equation}
r_{\alpha\beta}=\frac{\varsigma}{\varsigma_{max}}
\end{equation}
is obtained, which can be validated on empirical networks.

\begin{center}
\begin{table*}
\caption{\label{tab:1} Different types of multilayer networks.}
\begin{tabular}{l|p{8cm}|ll}\hline \hline
Term &  Explanation &  Mathematical definition  &\\
\hline
Multilayer network & Generic term for networks with multiple layers.  &     &\\ \hline
Multiplex network & Each layer has the same set of nodes (or an overlap of a fraction nodes) but different links among them.  &  $V_{\alpha}\cap V_{\beta}=V_{M}=V$, $\forall\alpha\ne\beta$   &\\ \hline
Interdependent network & Each layer typically has different types of nodes, and there are dependency links (not physical connections) between the nodes in different layers.  &  $V_{\alpha}\cap V_{\beta}=\emptyset$, $\forall\alpha\ne\beta$   &\\ \hline
Interconnected network & Each layer typically has different types of nodes, and there are actual physical links between the nodes in different layers.  &  $V_{\alpha}\cap V_{\beta}=\emptyset$, $\forall\alpha\ne\beta$     &\\
\hline \hline
\end{tabular}
\end{table*}
\end{center}

\subsection{The classification of multilayer networks}
As mentioned before, the term ``multilayer'' seems to have originated already in the late 1930s. Here, we use it as a proxy for various types of networks that are in one way or another formed by more than just a single, isolated network. A few examples are shown in Fig.~\ref{illustrate2}. By using ``multilayer network'' as the most general term, we are following the convention in the field, where for example the robustness of multilayer networks has recently received notable attention \cite{zhou_d_pre13, peixoto2012evolution, baxter2012avalanche}. In general, however, quite a variety of different network concepts, sometimes using just different names but studying the same thing, has emerged over the years. Examples include multiplex networks \cite{sola2013eigenvector, zhao2014immunization, battiston2014structural, nicosia2014measuring}, temporal networks \cite{holme2012temporal, masuda2013predicting, perra2012activity}, interconnected networks \cite{wang2013effect, aguirre2014synchronization, donges2011investigating, saumell2012epidemic}, multivariate networks \cite{pattison1999logit}, multidimensional networks \cite{berlingerio2013multidimensional}, cognitive social structures \cite{krackhardt1987cognitive}, as well as interdependent networks \cite{parshani_prl10, huang_xq_pre11, gao_jx_np12, li_w_prl12, dong_g_epl13}. In the following subsections, we provide a concise overview of some of these network concepts, although we primarily focus on those that have thus far been considered in the context of evolutionary games. A summary is provided in Table~\ref{tab:1}.

\subsubsection{Multiplex networks}
In a multiplex network all the layers contain the same set of nodes or share at least some fraction of the nodes. The difference between the layers is the way the nodes are connected with each other in each particular layer. Archetypical examples of multiplex networks are to be found in social and engineering systems. For example, one can construct a multiplex network from the scientific collaboration network and the citation network \cite{li2012har}. The network of airports can be translated into a multiplex form with different layers consisting of the routes of different airplane carriers \cite{cardillo2012modeling, nicosia2014measuring}. Other well-known examples include online social networks \cite{halu2013multiplex}, biological metabolic networks \cite{li2011integrative}, and road transportation networks \cite{cozzo2013clustering, rombach2014core}. If the set of nodes are the same in each layer (i.e., $V_{1}=V_{2}=\dots­=V_{M}=V$), the mathematical formulation of a multiplex network becomes $G_M=(V, E_M)$.

\subsubsection{Interdependent networks}
An interdependent network is typical made up of two or more different networks, such that there is little or no overlap between the nodes in the different layers. The wellbeing of nodes in a particular layer depend on the wellbeing of nodes in a different layer, and vice versa. There thus exist so-called dependency links between the nodes that are part of different layers. These links are not actual physical links, but rather imaginary links that denote the co-dependence; hence the name interdependent networks. The concept of interdependent networks was first proposed in the seminal paper by Buldyrev et al. \cite{buldyrev2010catastrophic}, where cascading failures between an electrical grid network and a computer network have been studied. Subsequently, the concept became very popular and used to study various other phenomena that might be affected by dependency links between different networks. The networks of airports and seaports can also be described as interdependent networks, because the proper functioning of an airport in a city depends on the resources that are deliver by sea, and similarly, the proper functioning of a seaport may depend on goods delivered by air \cite{parshani2010inter}. Moreover, food webs constructed from species which depend on other species are also interdependent when the same species participate in different webs \cite{pocock2012robustness}. Lastly, we note that since the availability of credit from the banking network and the economic production by the network of commercial firms are interdependent, an interdependent network model of banks and bank assets has also been used to analyze the propagation of failures in the economy \cite{huang2013cascading}.

\subsubsection{Interconnected networks}
An interconnected network is similar to an interdependent network in that it is typical made up of two or more different networks, such that there is little or no overlap between the nodes in the different layers. In the interconnected network, however, there are actual physical links that connect together the nodes from different layers (rather than dependency links that we have described for interdependent networks). Interconnected networks can thus be regarded as interconnected communities or clusters within a single, larger network. Based on this theoretical framework, for example, the climate network can be decomposed into different network layers to exploit the stratification and circulation of the terrestrial atmosphere \cite{donges2011investigating}.

Summing up the classification of multilayer networks, we refer again to Fig.~\ref{illustrate2}, where we show schematic illustrations and empirical observations of above mentioned networks, and to Table~\ref{tab:1}, where we summarize the basic properties of these networks. In keeping with the main theme of this colloquium, we also now briefly touch upon the importance of different types of multilayer networks for the consideration of evolutionary games. The definition of an evolutionary game primarily entails strategies and payoffs, and both are susceptible to the formalism of a multilayer network. If evolutionary games are played on multiplex networks, strategy imitation and payoff accumulation can take place either in the local neighborhood of a particular layer or across the layers, since most of the nodes exist in all layer \cite{wang2014degree, gomez2012evolution}. On the other hand, when considering evolutionary games on interdependent networks, which typically contain different nodes in each layer, direct strategy exchange across the layers is not allowed. Rather, players in a given layer can obtain information concerning strategy and payoffs in the other layer via dependency links \cite{jin_q_srep14, wang_b_jsm12, wang_z_epl12, wang_z_srep13, szolnoki_njp13, wang_z_srep13b, wang_z_njp14, wang_z_jtb14}. Indeed, evolutionary games on interdependent network have to date received the most attention in comparison to evolutionary games on other multilayer networks. With regards to evolutionary games on interconnected networks, strategy can be imitated across the different layers, with the main difference to single isolated networks being that connections among the different layers are sparse in comparison to connections within each layer \cite{tang_cb_pone14, gomez-gardenes_pre12, jiang2013spreading} (which is of course the same definition as is commonly used for the identification of communities in networks \cite{fortunato}). In Section III, we will expand on the subject of evolutionary games on multilayer networks in detail. Beforehand, in the next subsection, we briefly provide references to works where different algorithms for the generation of multilayer networks have been presented.

\subsection{Algorithms for the generation of multilayer networks}
Similarly as for single-layer networks \cite{erd6s1960evolution, watts_dj_n98, barabasi_s99}, several algorithms have been proposed for the generation of multilayer networks \cite{nicosia2013growing, criado2012mathematical, de2013centrality, lee2012correlated, funk2010interacting, cardillo2013emergence}. Probably the simplest and the most straightforward approach is to first construct individual networks based on the traditional algorithms \cite{boccaletti_pr06,albert_rmp02,newman_siamr03}, and then subsequently to insert inter-layer connections among the existing networks based on some specific requirements, like degree-degree correlation \cite{funk2010interacting, min2013network, marceau2011modeling}. Somewhat less straightforward algorithms exploit network growth or modifications of static network models.

Algorithms that rely on network growth to arrive at the desired multilayer structure obviously require the number of nodes to increase as a function of time \cite{nicosia2013growing, kim2013coevolution}. Similarly to the growth of single-layer networks \cite{boccaletti_pr06, albert_rmp02, newman_siamr03}, preferential attachment rules appear to attract the most interest. For example, it has been suggested that the probability of placing an inter-layer connection is proportional to the intra-layer degree of all the related nodes in each layer \cite{nicosia2013non}. Along this line, further generalizations are possible to arrive at multilayer networks with very specific properties \cite{kivela_jcn14, boccaletti_pr14, fotouhi2014non}.

As mentioned before, an alternative is to consider modifications of static network models, where one needs to know the specific structural properties, such as the degree distribution or the degree correlation, of each layer in advance \cite{pattison1999logit, bianconi2013statistical, funk2010interacting, marceau2011modeling}. Subsequently, the multilayer architecture can be realized through the adjustment of the corresponding parameters within or across the layers (like in the configuration models for single-layer networks) \cite{menichetti2014weighted, wang2013exponential}.

\section{Evolutionary games on multilayer networks}
In line with the main concept of multilayer networks, evolutionary games staged on them entail players that occupy the nodes on different layers and interact with their neighbors within and between these layers. As in traditional single-network game theoretical models, players pass their strategies or adopt them from more successful competitors \cite{szabo_pr07,perc_bs10}, typically from within the same layer. Moreover, players may interact with other players from a different network layer, but strategy transfer between the layers is typically not considered permissible. This restriction is in place because otherwise the whole setup becomes practically equivalent to a single-layer network, with possibly a complicated interaction topology. The consideration is the same as with the formulation of interconnected networks, which in principle can always be reduced to a single-layer network. Accordingly, evolutionary games on interconnected networks are not of particular interest, as it is unlikely that mechanisms different from the ones already observed on isolated networks would govern the evolutionary process. The main point of interest is to determine and understand how multiplexity and various forms of interdependence among the individual network layers affect the previously-observed cooperation supporting mechanisms. Do these mechanisms remain valid or become irrelevant? Can we detect additional effects which further strengthen the already known mechanisms?

The first step is to clarify the possible consequences of different kinds of interactions between players who are staying in different layers. Since strategy evolution is based primarily on the payoff (utility) differences between players, the first choice could be to assume that a player can also collect payoffs from an external source. This practically means the payoff of a player in a given layer depends also on the state of a player or players in other layers. The next subsection expands on this option.

\subsection{Coupling through utilities}

By considering that a player's payoff depends also on the state of a player in another layer, there are still plenty of options to consider on how precisely such an interdependence could be formulated. The first main direction to follow could be to assume that the success, hence the payoff, of the external player will directly modify the payoff of our focal player. This assumption was in fact considered in \cite{wang_b_jsm12, tang_cb_pone14, gomez-gardenes_pre12, jiang2013spreading, santos_md_srep14}. The second option to consider could be that the actual strategy of the player in the other layer will contribute to the payoff of the focal player via a particular payoff matrix. In the latter case, the strategy of the player in the other layer plays a similar role to the strategy of a neighbor within the same layer. The only difference is that the strategy invasion is only possible within a particular layer, but not between the network layers.

Staying with the option that the payoff of the external player will directly modify the payoff of the focal player, we review the model and results in \cite{wang_z_epl12}, where the evolution of cooperation was first studied in a system where players were distributed on two interdependent networks. The above described connection between player $x$ and his external partner $x^\prime$ can be given as
\begin{equation}
U_x = \Phi P_x + (1-\Phi) P_{x^\prime}\,,\, U_{x^\prime} = (1-\Phi) P_{x^\prime} + \Phi P_x\,,
\label{utility}
\end{equation}
where $\Phi$ determines the bias in the consideration of payoffs collected by the corresponding players $x$ and $x^\prime$ in the two networks.

It has been shown that the stronger the bias in the utility function, the higher the level of public cooperation. Due to the symmetry breaking, unequal levels of cooperation can be observed on the two layers, yet still, the aggregate density of cooperators on both networks is higher than the one attainable on an isolated network. This positive effect of biased utility functions is due to the suppressed feedback of individual success, which leads to a spontaneous separation of characteristic time scales of the evolutionary process on the two interdependent networks. Consequently, cooperation is promoted indirectly because the aggressive invasion of defectors is more sensitive to the slowing-down than the careful build-up of collective efforts in sizable groups.

\begin{figure}
\centerline{\epsfig{file=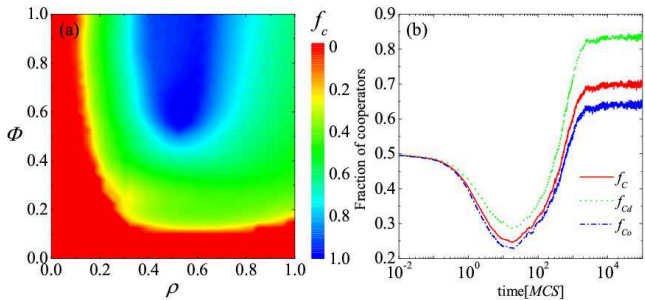,width=8.5cm}}
\caption{Panel (a) shows  the color coded fraction of cooperators $f_C$ in dependence on the fraction of players $\rho$ that are allowed to form an external link and the strength of these links $\Phi$, as obtained for the temptation to defect $b=1.05$ in the prisoner's dilemma game. Panel (b) shows that distinguished players, who have an external link with their
corresponding player in the other network, are more likely to cooperate
than those who do not have a link to the other layer. This follows from the time evolution of the fraction of cooperators in the whole population ($f_C$), among the distinguished players ($f_{C_d}$), and among ordinary players ($f_{C_o}$). It can be observed that
$f_{C_d}> f_{C_o}$. Parameter values used were: $b=1.05, \rho=0.3$ and $\Phi=0.8$. For details see \cite{wang_z_srep13b}.}
\label{optimal}
\end{figure}

One may argue that the introduced asymmetry via payoff links directly supports cooperators because it reveals the difference of speeds of spreading between the two competing strategies. But this is not necessarily true because the positive contribution of interdependent topology to the fundamental effect of network reciprocity can also be observed if the coupling is symmetric \cite{wang_z_srep13}. The phenomenon was referred to as spontaneous emergence of interdependent network reciprocity, which has proven to be extremely effective for maintaining considerable cooperation levels even at extremely adverse conditions. The key mechanism here is a simultaneous formation of correlated cooperator clusters on both networks. In the absence of this, when such a coordination process is disturbed, network reciprocity fails on both networks, leaving an undesired outcome in the whole system.

\begin{figure}
\centerline{\epsfig{file=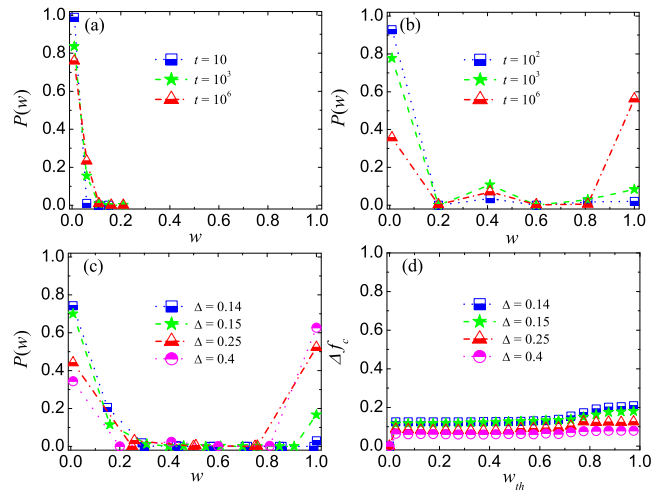,width=8.5cm}}
\caption{The spontaneous emergence of a two-class society when coevolution of interdependency is allowed. Panel (a) illustrates that the initially homogeneous teaching activity distribution remains practically untouched if the reward (punishment) parameter $\Delta$ is low.  For intermediate $\Delta=0.4$, however, practically a two-peak distribution evolves in the long-time limit, as shown in panel (b). This cooperator supporting distribution emerges for a broad range of control parameter, as evidence in panel (c). Lastly, panel (d) shows the excess cooperation level in the upper class society in relation to the average level in the whole system, suggesting that cooperation is more likely among those who are able to control their neighborhoods. For details see \cite{wang_z_njp14}.}
\label{two_class}
\end{figure}

Until this point, we have assumed that every player has an external link to the other layer and can thus collect an additional payoff to the one attainable within the home network. But this assumption is of course not always realistic, as not everybody is keen on maintaining such external links. It can thus easily happen that there is a kind of heterogeneity among the players, where just some specific fraction of them has an external link established to the other layer. According to previous observations, an intuitive expectation would be that it is better if all players have such a link to strengthen the above mentioned interdependent reciprocity. Hence, the prediction could be ``the more (external links) is better''. But the reality is in fact different. It was demonstrated that there is actually an optional fraction of distinguished players $\rho$, whose external links will provide the most effective interdependence between subnetworks for the successful evolution of cooperation \cite{wang_z_srep13b}. This phenomenon is illustrated in Fig.~\ref{optimal}. It can thus be concluded that there is an optimal interdependence which warrants the highest cooperation level. The explanation of this behavior is based on a phenomenon which is generally valid also on a single network. More precisely, it is better if there is an inhomogeneity among players because it helps forming homogeneous groups in the population \cite{santos_prl05, szolnoki_epl07, perc_pre08, szolnoki_njp08}. However, the formation of uniform patches is beneficial for cooperators because it reveals the advantage of mutual cooperation against defection. As a consequence, we can find more cooperators among those who influence others or among those who determine the strategy choice within their neighborhood. This reasoning is valid for interdependent networks just as much as it is valid for single-layer networks, as demonstrated in the right panel of Fig.~\ref{optimal}. The left panel of Fig.~\ref{optimal} illustrates in addition that a much higher cooperation level can be reached on interdependent networks than is attainable on an isolated network (which corresponds to the $\rho=0$ case). Research thus shows that interdependent networks are likely to augment those cooperator-supporting mechanisms that have already been observed previously single-layer, traditional network.

\begin{figure}
\centerline{\epsfig{file=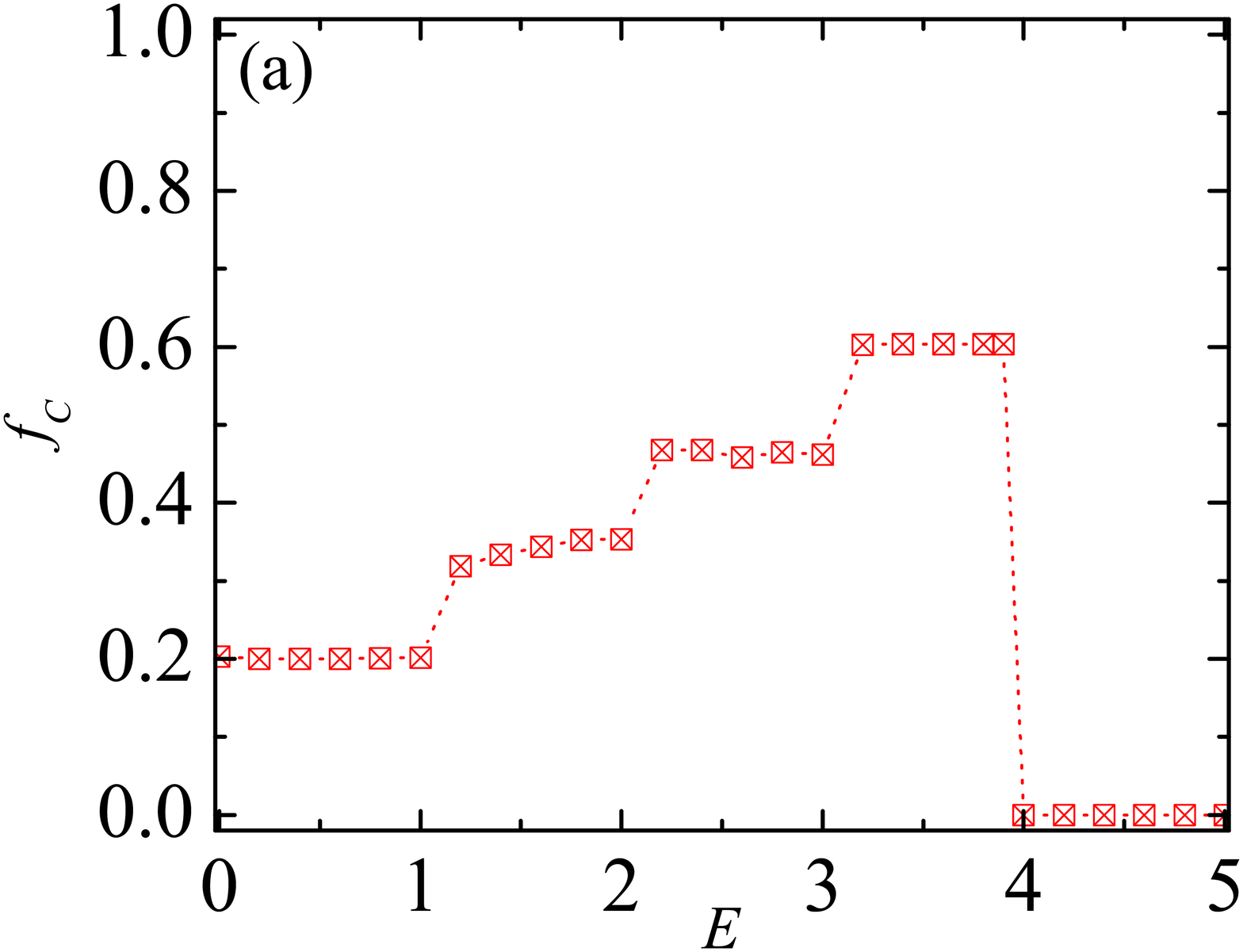,width=4.25cm}\epsfig{file=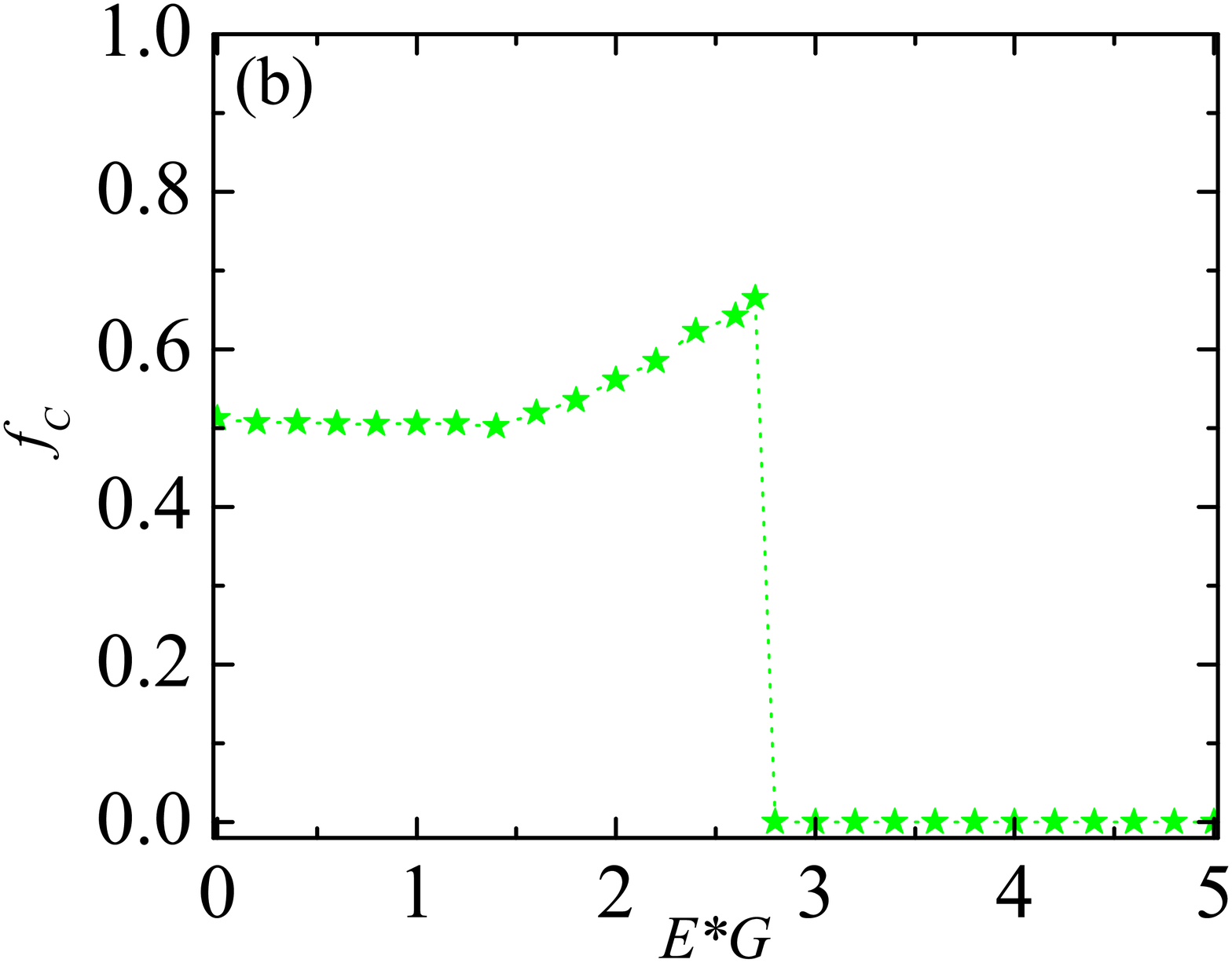,width=4.25cm}}
\caption{Intermediate values of the utility threshold $E$ are able to sustain widespread cooperation both in the prisoner's dilemma (a) and in the public goods (b) game. For both cases each layer is formed by a square lattice where the group size is $G=5$. For details see \cite{wang_z_jtb14}.}
\label{rewarding}
\end{figure}

The next logical step could be to find out whether such an optimal interdependency can emerge spontaneously. To explore this option, a coevolutionary rule between strategy change and network interdependence has been proposed and studied in \cite{wang_z_njp14}. More precisely, it was assumed that a successful player, who can pass his strategy to a neighbor, is awarded by increasing its individual teaching activity (or strategy pass capacity) by a finite value $w=w+\Delta$. We note that this property of the donor player is incorporated in the applied Fermi strategy adoption probability via a pre-factor $w_x$, namely $\Gamma(s_{x} \rightarrow s_{y})= w_x ({1+\exp[(U_y-U_x)/K]})^{-1}$. Importantly, if the attempted strategy pass was unsuccessful, the teaching activity was reduced by the same value. Furthermore, if the strategy pass capacity $w_x$ exceeded a threshold value, then the corresponding player was granted an external link towards the other layer. By applying this protocol, it was demonstrated that an optimal interdependence between graphs can spontaneously evolve which can maintain cooperation even under extremely adverse conditions. The explanation of this effective construction is based on the spontaneous emergence of a two-class society where only the upper class is being allowed to control and take advantage of the interdependence. This segregation is illustrated in Fig.~\ref{two_class}. Based on the previously described argument involving the inhomogeneity of players, it is a natural consequence that cooperative players are more competent in sustaining compact clusters of followers if they reach the upper class. To conclude, the impact of interdependence between networks can thus be exploited successfully only by cooperators, which in turn extends parameter intervals where these seemingly weaker competitors are able to survive.

The importance of details that determine coupling between otherwise independent networks can be illustrated nicely in a biology motivated model \cite{wang_z_jtb14}. Here the relation between the fitness and external demands, denoted by a threshold value $E$ , has a decisive role. According to the suggested model, initially all players belong to one independent structured population. Simultaneously with the strategy evolution, players whose current utility exceeds a threshold are rewarded by an external link to a player belonging to the other population. Driven by the same motivation, as soon as the utility drops below the threshold, the external link is terminated. In this way, the individual fitness of a player and its chance of having an external connection are strongly correlated. As a consequence, a time-varying interdependence evolves between the networks. It turned out that, regardless of the details of the evolutionary game and the interaction structure, the self-organization of fitness and reward gives rise to distinguished players that act as strong catalysts of cooperative behavior. However, if the utility thresholds value $E$ is too large distinguished players are no longer able to percolate \cite{wang_z_pre12b, wang_z_srep12}.
Hence the interdependence between the two populations vanishes, and cooperators are forced to rely on traditional network reciprocity alone, which generally is able to sustain a lower level of cooperative behavior. This threshold dependence of the cooperation level is illustrated in Fig.~\ref{rewarding}. It is worth noting that a similar process, namely the formation of links outside the immediate community, seems particularly applicable also in human societies, where an individual is typically member in many different social networks, both in real life, as well as online on networks such as Facebook or Twitter.

\begin{figure}
\centerline{\epsfig{file=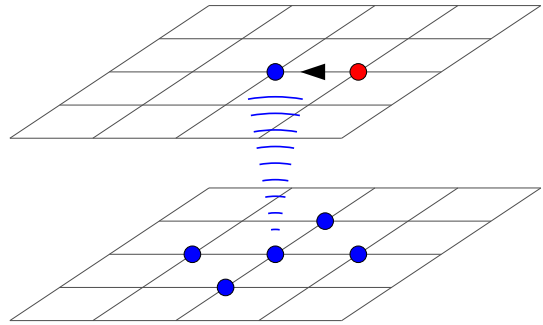,width=7.5cm}}
\caption{Information sharing between networks affects strategy transfer
between neighboring players. The red player in the upper network tries to transfer its strategy to the blue player. Meanwhile, the blue player receives information from the corresponding players in the bottom network that they all adopt the same (blue) strategy. Because of this, the blue player in the upper network is reluctant to change its strategy to red, despite of the fact that the red player might have a higher payoff.}
\label{setup}
\end{figure}

\begin{figure*}
\centerline{\epsfig{file=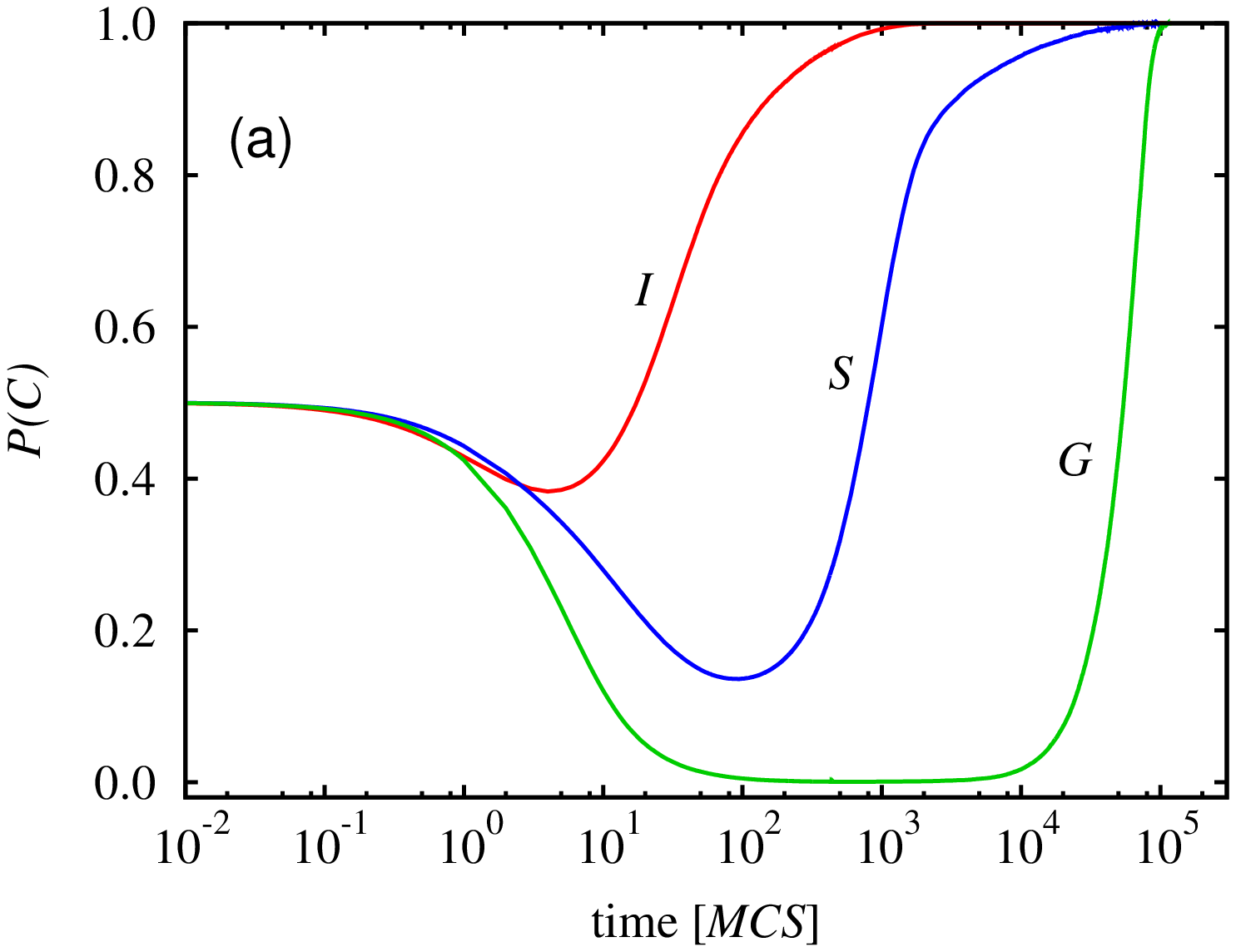,width=7.2cm}\epsfig{file=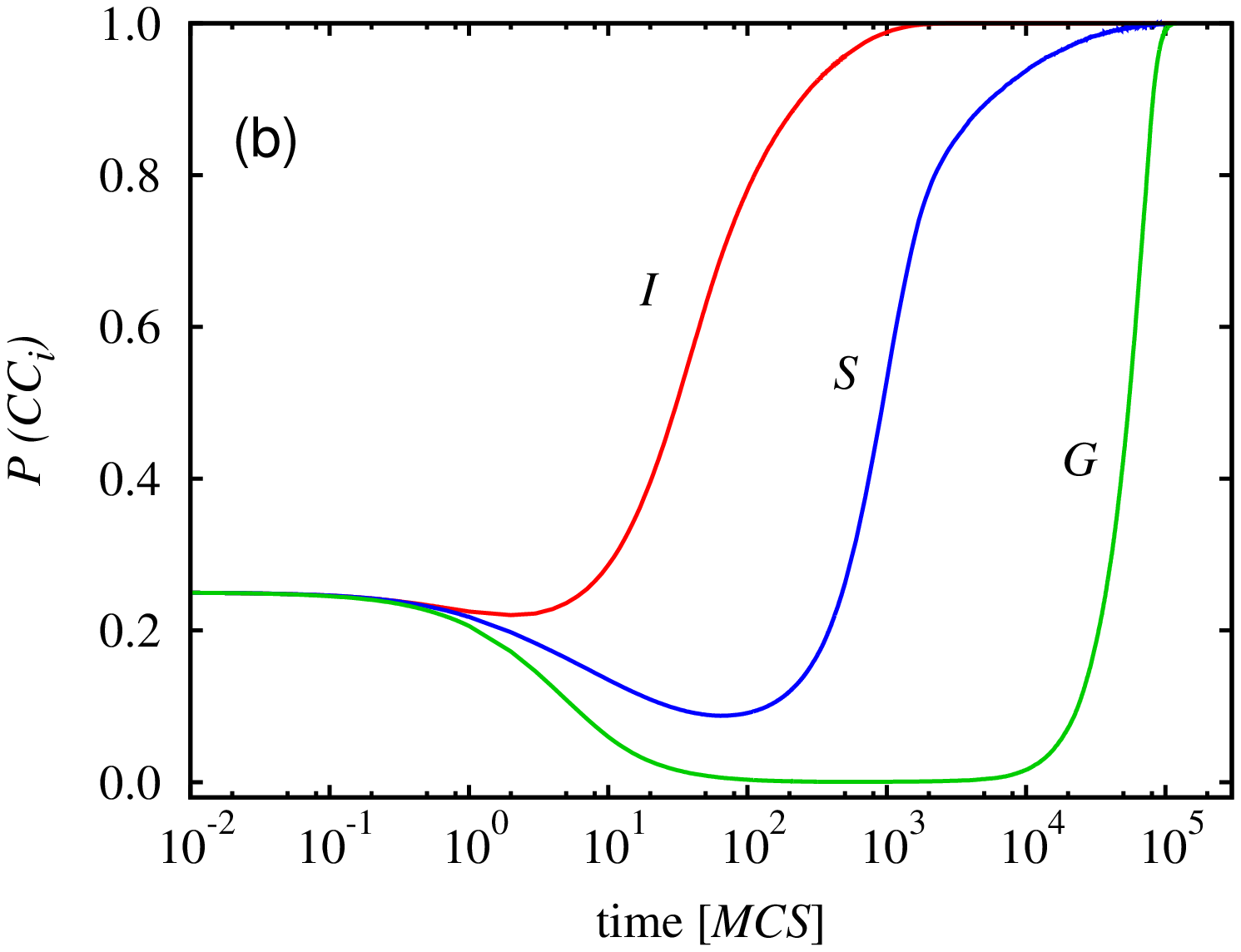,width=7.2cm}}
\centerline{\epsfig{file=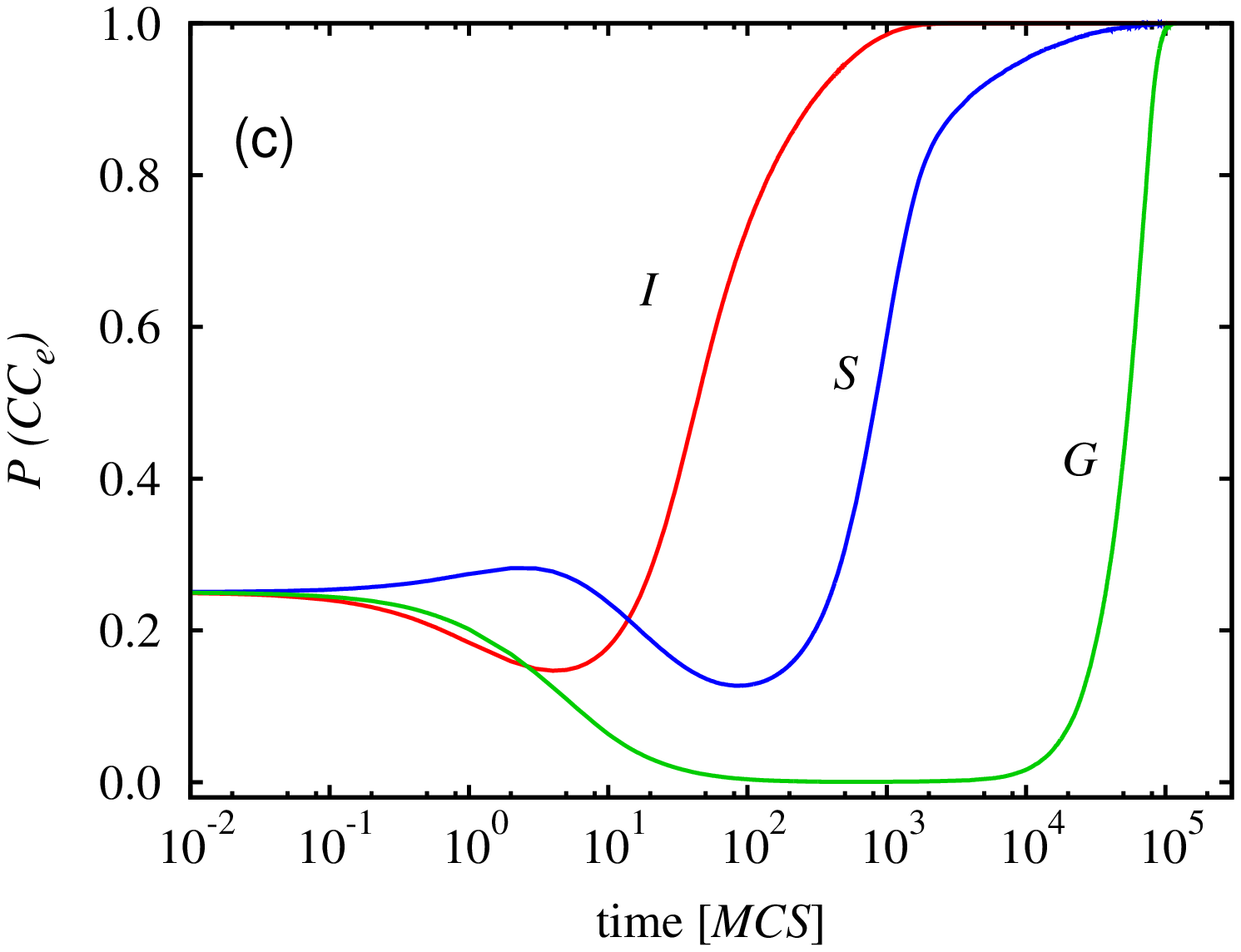,width=7.2cm}\epsfig{file=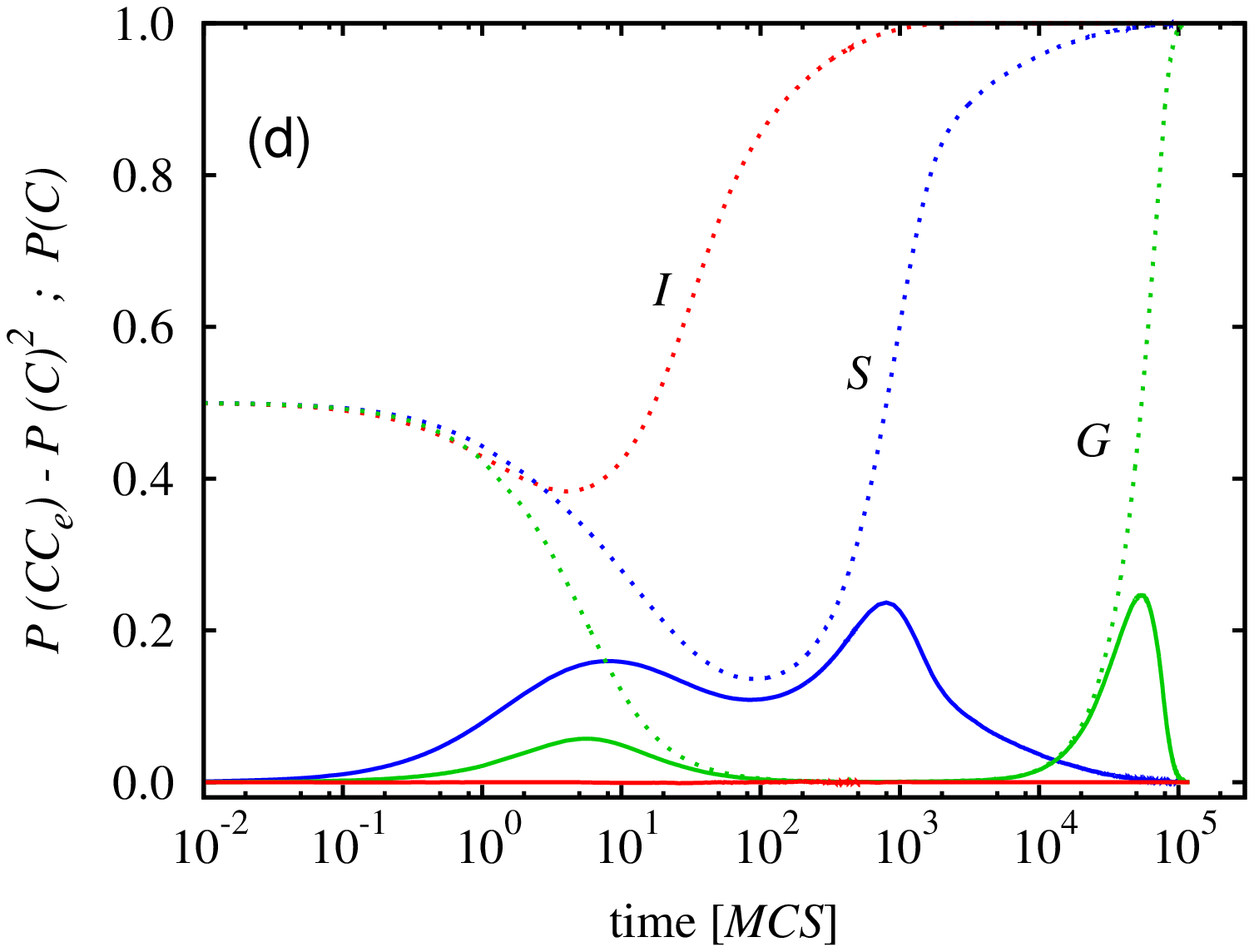,width=7.2cm}}
\caption{Time evolution of cooperation and cooperative pairs within and between networks reveals why information sharing promotes cooperation. Panel (a) shows the probability of finding a cooperator $P(C)$ in dependence on time, panel (b) shows the probability of finding $C-C$ pairs within a network $P(C C_i)$, while panel (c) shows the probability of finding $C-C$ pairs between the two networks $P(C C_e)$. Solid lines in panel (d) depict the excess correlation of cooperators between the two networks, determined as $P(C C_e) - P(C)^2$. For easier reference, panel (d) also features $P(C)$ as dotted lines. For details see \cite{szolnoki_njp13}.}
\label{time_evol}
\end{figure*}

The work presented in \cite{jin_q_srep14} reveals another interesting result. Namely, if the symmetry constrain of individual utility functions between two populations remains in tact (otherwise being similar to Eq.~\ref{utility}), then there exists a threshold value of coupling  $\Phi_C$ that leads to more favorable conditions for the evolution of cooperation. Below the critical value, the cooperation level is equal on both layers and increases monotonously as the coupling strength $\Phi$ increases. At variance, if the utility correlation between the two interdependent networks exceeds the $\Phi_C$ threshold, a spontaneous symmetry breaking between cooperation levels on the two networks emerges, and this irrespective of the details of the interaction topology. Moreover, it has been found that the final state is closely related to the evolution of heterogeneous strategy pairs across network layers. The asymmetric expansion of heterogeneous strategy pairs namely plays a pivotal role in the symmetry breaking process. This in turn enhances traditional network reciprocity \cite{nowak_n92b, szabo_pr07}. Along this line, evolutionary games based on multi-point interactions and with more than two competing strategies \cite{perc_jrsi13}, like the public goods game with volunteering, might be worth considering in the future.

\subsection{Alternative ways of coupling}
Although payoffs play an all important and prominent role in the success of strategies in the realm of evolutionary game theory, there are nevertheless alternative ways to the ones reviewed thus far for establishing a connection between two or more otherwise independent (isolated) networks (populations).

We may assume, for instance, that a player's decision to adopt a particular strategy is based not only on the payoffs of its neighbors in a given network, but also on the popularity of the potential new strategy in another network \cite{szolnoki_njp13}. A schematic example to illustrate this point is shown in Fig.~\ref{setup}, where the willingness to adopt a more successful new strategy is significantly lowered if the old strategy is more frequent in the neighborhood of the corresponding player who resides in a different network. Naturally, the reversed situation applies too. Namely, if the potential new strategy is popular in the other network, then this should amplify the likelihood of its acceptance in the current network.

\begin{figure*}
\centerline{\epsfig{file=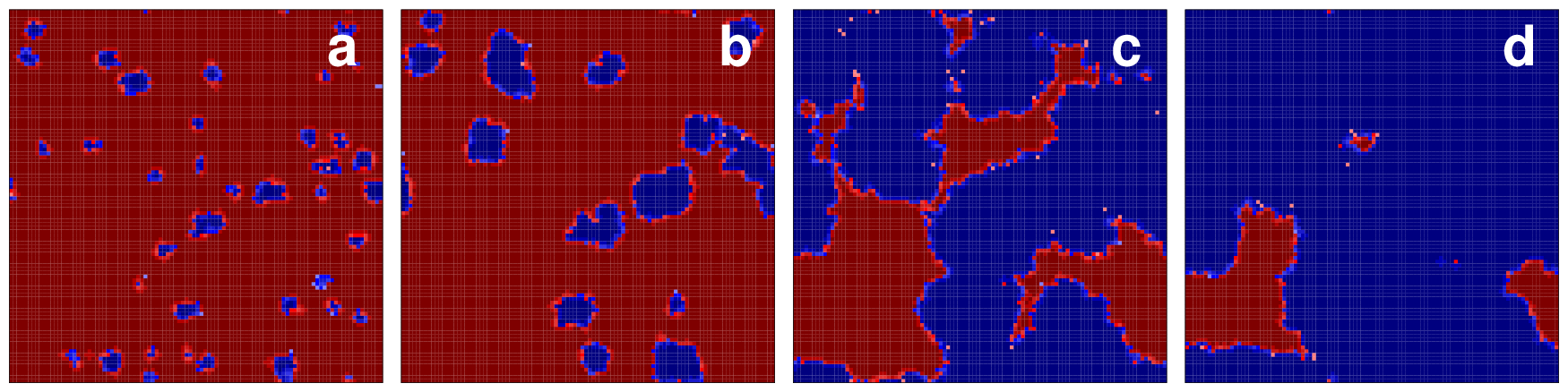,width=15cm}}
\centerline{\epsfig{file=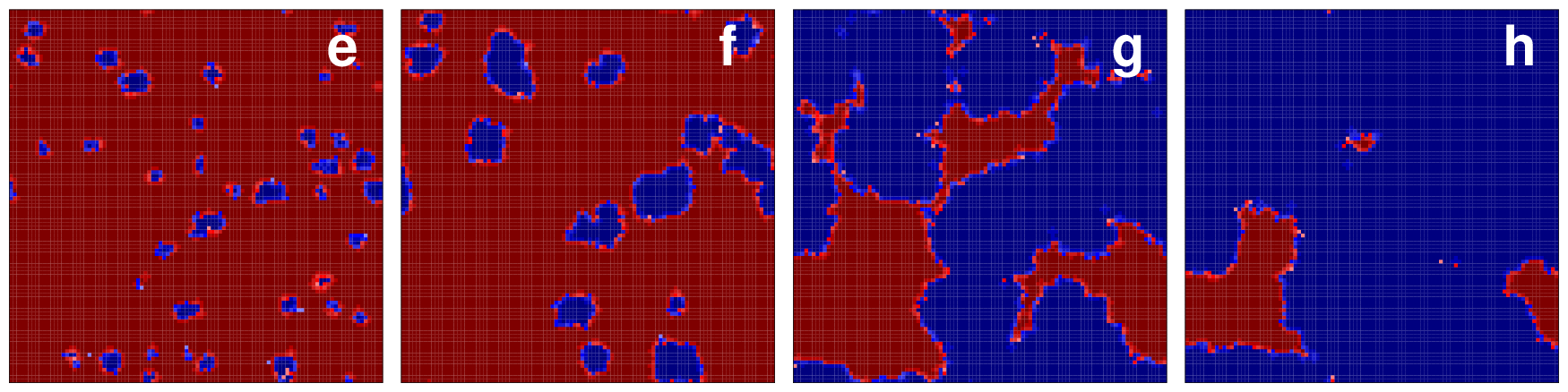,width=15cm}}
\caption{Characteristic snapshots reveal the spontaneous emergence of strongly correlated evolution that is due to information sharing. Presented are snapshots of the upper (a)-(d) and the lower (e)-(h) network layer, as obtained with the information-transfer model presented in \cite{szolnoki_njp13}. Defectors are denoted by red and cooperators are denoted by blue. The random initial state and the final pure $C$ phase are not shown. For further details see \cite{szolnoki_njp13}.}
\label{snapshots}
\end{figure*}

Despite of the fact that the proposed protocol is strategy neutral, this kind of coupling generates an environment which is beneficial for the evolution of cooperation. As a result, cooperators can survive in parameter regions which would produce an all $D$ phase if the game would be staged in a single-layer network. The mechanism which is responsible for this improvement is based on the spontaneous emergence of synchronized strategy evolution on different networks. As Fig.~\ref{time_evol} demonstrates, there is an enhanced correlation between the two networks in terms of how the strategies evolve. Interestingly, such a coordination has different consequence for cooperators than it has for defectors. In particular, while it supports the stability of compact cooperative domains, it also simultaneously slows down the propagation of defection. This spectacular synchronized evolution is illustrated in Fig.~\ref{snapshots}, where the spatial distribution of strategies is illustrated separately for the upper and lower network at the same times from left to right. Importantly, we note that there is no payoff-driven coupling between the two networks.

An interesting alternative way of coupling, not based on player payoffs, has also been proposed by Lugo and San Miguel \cite{lugo2015learning}, who considered a two-layer network where each individual is connected to a so-called ``playing'' and ``learning'' network . It was observed that the degree of social pressure via the level of doubt, i.e., the skepticism related to the wisdom of crowd \cite{szolnoki_srep12, szolnoki_rsif15}, may play a decisive role in the evolution of cooperation.

We close this section by mentioning yet another independent way on how interdependent networks may arise in the framework of evolutionary game theory. In contrast to the previously discussed cases, where players formed independent populations which were then coupled in a specific way, it is also possible to consider interdependent networks where the same players are members on different networks simultaneously. However, in line with the classification of multilayer networks presented in Section II, this approach is thus most accurately classified as evolutionary games on multiplex networks, which was also the terminology adopted G{\'o}mez-Garde{\~n}es et al. \cite{gomez-gardenes_srep12}, who conducted the original research. The study revealed that, if certain players are members in more than one network, before unseen new phenomena may emerge that additionally favor prosocial behavior. For example, the enhanced resilience of cooperation can be due to a non-trivial organization of cooperation across different network layers. Accordingly, we note that this class of models, namely evolutionary games on multiplex networks (for earlier related work see \cite{gomez-gardenes_c11, gomez-gardenes_epl11}), certainly provides a promising research avenue to explore in the future.

\section{Conclusions and outlook}
This colloquium on evolutionary games on multilayer networks is motivated by the fact that networks of networks are often a significantly more apt description of real-life systems than isolated networks, and of course also more apt than well-mixed models \cite{buldyrev2010catastrophic, gao_jx_nsr14, helbing_n13, kivela_jcn14, boccaletti_pr14, d2014networks}. The latter two approaches are invaluable for conducting proof-of-principle fundamental research, yet when it comes to actual models that ought to address a more specific facet of reality, especially in human societies, then networks of networks are hardly avoidable. This is why the efforts to clarify the consequences of interdependence for the outcome of evolutionary games should not be dismissed as being purely ``academic'', but rather acknowledge the fact that they bring our models a step closer to real life problems. While it is straightforward to realize that the reviewed theoretical predictions will be difficult to test empirically by means of human or economic experiments, we argue that this adds to the relevance of theoretical research and mathematical modeling as basically the only means through which we may hope to obtain quantitative insights into the relevance of multilayer networks for the outcome of evolutionary games.

From the accumulated theoretical knowledge, we may once filter out the most relevant aspects of human interactions that drive our cooperative behavior. This perspective is everything but a far-fetched illusion, as recent research clearly underlines and emphasizes the utility and relevance of simulations and mathematical modeling, for example as a key tool for outbreak response \cite{lofgren2014opinion} as well as for vitally informing research in evolutionary biology \cite{servedio2014not}. Beyond interactions among living organisms, one also does not need to look far to discover several application points in social management systems, where models are often more accurate if we assume the overall interaction topology to be somehow linked but otherwise separate networks \cite{cardillo2012modeling, berlingerio2013multidimensional}.

We hope that the content of this review will be a useful source of information, both in terms of the basic concepts and definitions that pertain to multilayer networks, and even more so in terms of the beautiful perspective that is offered by evolutionary games on multilayer networks. The take home message is that several mechanisms have already been discovered by means of which the interdependence between different networks or network layers may help to resolve social dilemmas beyond the potency of traditional network reciprocity \cite{nowak_n92b}. A prominent example is interdependent network reciprocity \cite{wang_z_srep13}, which is capable to maintain healthy levels of public cooperation even under extremely adverse conditions. Network interdependence can thus be exploited effectively to promote cooperation past the limits imposed by isolated networks, but only if the coordination between the interdependent networks is not disturbed. Other mechanisms that promote the evolution of cooperation and build prominently on networks of networks include non-trivial organization of cooperators across the interdependent network layers \cite{gomez-gardenes_srep12}, probabilistic interconnectedness \cite{wang_b_jsm12}, information transmission between different networks \cite{szolnoki_njp13}, rewarding evolutionary fitness by enabling links between populations \cite{jiang2013spreading, wang_z_jtb14}, as well as self-organization towards optimally interdependent networks by means of coevolution \cite{wang_z_njp14}, all of which we have reviewed in this colloquium.

Directions for future research are many, and the outlook for joining the field is thus promising. In terms of evolutionary games, perhaps the most obvious path to take is considering other types of games on networks of networks. Viable candidates include the ultimatum game \cite{guth1982experimental, page_prsb00, kuperman_epjb08, gao_j_epl11, szolnoki_prl12, wu_t_srep13}, rock-paper-scissors games \cite{szolnoki_jrsif14}, the naming game \cite{baronchelli2006topology}, or the collective-risk social dilemma game \cite{pacheco_plrev14, milinski2008collective, santos_pnas11, chen_xj_epl12, vasconcelos_ncc13}. Here the focus is frequently on the emergence of pattern formation and collective behavior such as fairness, species diversity, cyclical dominance, language evolution, or the prevention of dangerous climate change \cite{pacheco_plrev14, szolnoki_plrev14}. In addition to these options, disease-spreading processes can also been embedded into the concept of multilayer networks. For example, contagion spreads on one layer while prevention measures, like vaccination, are supported by another layer \cite{bauch2013epidemiology}. As is well-known, game theoretical models are frequently employed to study the impact of different strategic choices on disease prevention in isolated networks \cite{bauch_pnas04, d-onofrio_jtb10, fu2011imitation}. Here the multilayer theoretical framework, paired with evolutionary game theory, provides a fascinating gateway towards richer and more detailed epidemiology research. Lastly, informed by the interconnectedness of different means of transport, the consideration of game theoretical models in multilayer transport frameworks also promises interesting discoveries for behavioral traffic research \cite{gu2011onset}, especially when combined with the optimization of transport costs and the transition efficiency in empirical networks.

We conclude with the hope that our colloquium will be motivational towards the consideration of at least some of the above research avenues in the near future.

\begin{acknowledgments}
This work was supported by the National Natural Science Foundation of China (Grant No. 61374169), the Hungarian National Research Fund (Grant K-101490), the Slovenian Research Agency (Grant P5-0027), and by the Deanship of Scientific Research, King Abdulaziz University (Grant 76-130-35-HiCi).
\end{acknowledgments}

\end{document}